\input{aipcheck}

\documentclass[
    ,final            
  ]
  {aipproc}

\layoutstyle{6x9}

\begin{document}

\title{Supernovae as Probes of Extra Dimensions}


\author{V. H. Satheesh Kumar} {address={School of Physics, University of Hyderabad, Hyderabad 500 046, India.},altaddress={{Department of Physics, Jain International Residential School, Bangalore 562 112, India.}}} 
\author{P. K. Suresh}{
  address={School of Physics, University of Hyderabad, Hyderabad 500 046, India.}
}

\author{P. K. Das}{
  address={The Institute of Mathematical Sciences, CIT Campus, Taramani, Chennai 600 113, India.}
}

\begin{abstract}
Since the dawn of the new millennium, there has been a revived interest 
in the concept of extra dimensions.
 In this scenario all the standard  model matter and gauge fields are confined to the 4 dimensions and  only gravity can escape to higher dimensions of the universe.
This idea can be tested using table-top experiments, collider experiments, astrophysical or cosmological observations. The main astrophysical constraints come from the cooling rate of supernovae, neutron stars, red giants and the sun. In this article, we consider the energy loss mechanism of SN1987A and study the constraints it places on the number and size of extra dimensions and the higher dimensional Planck scale.
\end{abstract}

\maketitle


\section{Introduction}
It has been recently noted that the scale of quantum gravity $M_{Pl}\approx 10^{19}$ GeV can be brought down to a few TeV in certain class of extra dimensional models \cite{ADD}. The only experimentally verified scale $M_{EW}$ of Standard Model(SM) interactions in four dimensions lies within the TeV scale which allows tolerable quantum corrections. Therefore, the assumption that the 4+n dimensional gravity becomes strong at  TeV scale, while the standard gauge interactions remain confined to the four-dimensional spacetime does not conflict with today's data available from low energy gravitational experiments \cite{LCCGVP}. Such a notion of TeV scale gravity solves the hierarchy problem between $M_{EW}$ and $M_{Pl}$ without relying on supersymmetry or technicolour. According to this model, the observed weakness of gravity at long distances is due to the presence of $n$ new spatial dimensions large compared to the electroweak scale. This can be inferred from the relation between the Planck scales of the $D=4+n$ dimensional theory $M_{D}$ and the  four dimensional theory $M_{Pl}$, which is given by
\begin{equation}
M_{Pl}^2 \sim R^n M_{D}^{n+2},
\end{equation}
where $R$ is the size of the extra dimensions. Putting $M_{D} \sim 1$ TeV one finds
\begin{equation}
R \sim 10^{\frac{30}{n} - 17} \mbox{cm}.
\end{equation}
For $n=1$, $R \sim 10^{13}$ cm, which is obviously excluded since it would  modify the Newtonian gravity at solar-system distances. For $n=2$, we get $R \sim 1$ mm, which is precisely the distance where our present experimental measurement of gravitational strength stops. Clearly, while the gravitational force has not been directly measured below a millimeter, the success of the SM up to $\sim 100$ GeV implies that the SM fields can not feel these extra dimensions, that is they are confined to only  `3+1-brane', in the higher dimensional spacetime called `bulk'.
In this framework the universe is $4+n$  dimensional with the fundamental Planck scale $M_D$ near the weak scale, with $n \geq 2$ new sub-mm sized dimensions where gravity can freely propagate everywhere in the bulk, but the SM particles are localised on the 3-brane embedded in this bulk.
This theory predicts a variety of novel signals which can be tested using table-top experiments, collider experiments, astrophysical or cosmological observations. It has been pointed out that one of the strongest constraints on this physics comes from SN1987A~\cite{ADD2}. Various authors have done calculations to place such constraints on $M_D$ and $n$ \cite{Cullen:1999hc,BHKZ,Hanhart:2001er,Hanhart:2001fx,Hannestad:2001jv,Hannestad:2003yd,Malcolm}. In this article, we summarise all the results which have appeared in the literature so far.

\section{Supernova Explosion and Cooling}
Supernovae come in two main observational varieties \cite{VHS}. 
Those whose optical spectra exhibit hydrogen lines are classified as Type II, while hydrogen-deficient SNe are designated Type I. 
Physically, there are two fundamental types of supernovae, based on what mechanism powers them: the thermonuclear SNe and the core-collapse ones. Only SNe Ia are thermonuclear type and the rest are formed by core-collapse of a massive star. 

The core-collapse supernovae are the class of explosions which mark the evolutionary end of massive stars ($M \geq 8\,M_\odot$). 
 The collapse can not ignite nuclear fusion because iron is the most tightly bound nucleus. Therefore, the collapse continues until the equation of state stiffens by nucleon degeneracy pressure at about nuclear density ($3\times10^{14}~{\rm g~cm^{-3}}$).  At this ``bounce'' a shock wave forms, moving outward and expelling the stellar mantle and envelope.
 The kinetic energy of the explosion carries about 1\% of the liberated gravitational binding energy of about $3\times10^{53}~{\rm erg}$ and the remaining 99\% going into neutrinos. This powerful
and detectable neutrino burst is the main astro-particle interest of core-collapse SNe.

In the case of SN 1987A, about $10^{53}$ ergs of gravitational binding energy was released in few seconds and the neutrino fluxes were measured by Kamiokande \cite{Kamio} and IMB \cite{IMB} collaborations. Numerical neutrino light curves can be compared with the SN~1987A data where the measured energies are found to be ``too low''.  For example,
the numerical simulation in \cite{Totani:1997vj} yields time-integrated values $\langle E_{\nu_e}\rangle\approx13~{\rm MeV}$, $\langle E_{\bar\nu_e}\rangle\approx16~{\rm MeV}$, and $\langle E_{\nu_x}\rangle\approx23~{\rm MeV}$.  On the other hand, the data imply $\langle E_{\bar\nu_e}\rangle=7.5~{\rm MeV}$ at Kamiokande and 11.1~MeV at IMB~\cite{Jegerlehner:1996kx}.  Even the 95\% confidence range for Kamiokande implies $\langle E_{\bar\nu_e}\rangle<12~{\rm MeV}$.  Flavor oscillations would increase the expected energies and
thus enhance the discrepancy~\cite{Jegerlehner:1996kx}.  It has remained unclear if these and other anomalies of the SN~1987A neutrino signal should be blamed on small-number statistics, or point to a serious problem with the SN models or the detectors, or is there a new physics happening in SNe?

Since we have these measurements already at our disposal, now if we propose some novel channel through which the core of the supernova can lose energy, the luminosity in this channel should be low enough to preserve the agreement of neutrino observations with theory. That is,
${\cal L}_{new\, channel} \leq 10^{53}\, ergs\, s^{-1}.$
This idea was earlier used to put the strongest experimental upper bounds on the axion mass \cite{axions}. Here, we will consider  emission of higher-dimensional gravitons from the core. Once these particles are produced, they escape into  extra dimensions, carrying energy away with them. The constraint on luminosity of this process can be converted into a bound on the $M_D$.  The argument is very similar to that used to bound the axion-nucleon coupling strength \cite{axions, axionpapers,BBT,Raffelt}. The ``standard model" of supernovae does an exceptionally good job of predicting the duration and shape of the neutrino pulse from SN1987A. Any mechanism which leads to significant energy-loss from the core of the supernova immediately after bounce will produce a very different neutrino-pulse shape, and so will destroy this agreement
~\cite{BBT}. Raffelt has proposed a simple analytic criterion based on detailed supernova simulations~\cite{Raffelt}: if any energy-loss mechanism has an emissivity greater than $10^{19}$ ergs g$^{-1}$ s$^{-1}$ then it will remove sufficient energy from the explosion to invalidate the current understanding of Type-II supernovae's neutrino signal.

\section{Constraints on Extra Dimensions}
The most restrictive limits on $M_D$ come from SN~1987A energy-loss argument.  If large extra dimensions exist, the usual four dimensional graviton is complemented by a tower of Kaluza-Klein (KK) states, corresponding to new phase space in the bulk.   The KK gravitons interact with the strength of ordinary gravitons and thus are not trapped in the SN core. Each KK graviton state couples to the SM field with the usual gravitational strength according to \cite{Han:1999sg}
\begin{equation}
{\cal L}\ = -{\kappa\over2}\sum_{\vec n}
\int d^4 x\ h^{\mu\nu, {\vec n}} T_{\mu\nu}\ ,
\label{inter}
\end{equation}
where $\kappa^2 = 16 \pi G^{(4)}_N = \frac{16 \pi}{M^2_{Pl}}$ and the summation is over all KK states labeled by the level $\vec n$. $T_{\mu\nu}$ is the energy-momentum tensor of the SM fields residing on the brane and $h^{\mu\nu, {\vec n}}$ the KK state.
Since for large $R$ the KK gravitons are very light, they may be
copiously produced in high energy processes. For real emission of
the KK gravitons from a SM field, the total cross-section can be written as
\begin{equation}
\sigma_{\rm tot}\ =\ \kappa^2\sum_{\vec n} \sigma({\vec n})\ ,
\end{equation}
where the dependence on the gravitational coupling is factored out.
Because the mass separation of adjacent KK states,  ${\cal O}(1/R)$,
is usually much smaller than typical energies in a physical process,
we can approximate the summation by an integration.

Since we are concerned with the energy loss to gravitons escaping
into the extra dimensions, it is convenient and standard 
to define the quantities $\dot{\epsilon}_{a+b \to c}$
which are the rate at which energy is lost to gravitons via the
process $a + b \to c $, per unit time per unit mass of
the stellar object. In terms of the cross-section $\sigma_{a+b \to c}$
the number densities  $n_{a,b}$ for a,b and the mass density
$\rho$, $\dot{\epsilon}$ is given by
\begin{equation}
\dot{\epsilon}_{a + b \to c.} = \frac{\langle n_a n_b
\sigma_{(a+b \to c)} v_{rel} E_{c} \rangle}{\rho}
\label{emrate}
\end{equation}
where the brackets indicate thermal averaging.

 During the first few seconds after collapse, the core contains neutrons, protons, electrons, neutrinos and thermal photons. There are a number of processes in which KK gravitons can be produced. For the conditions that pertain in the core at this time ($T \sim 30-70$ MeV, $\rho \sim (3-10) \times 10^{14}$ g cm$^{-3}$), the relevant processes are
\begin{itemize}
\item  Nucleon-Nucleon Brehmstrahlung: $NN \to NNG_{KK}$
\item  Graviton production in photon fusion: $\gamma \gamma \to G_{KK}$,
and
\item  Electron-positron anhilation process: $e^{-} e^{+} \to G_{KK}$.
\end{itemize}
In the SNe, nucleon and photon abundances are comparable (actually nucleons are somewhat more abundant). Nucleon-nucleon bremhmstrahlung is the dominant process relevant for the SN1987A where the temperature is comparable to $m_\pi$ and so the strong interaction between N's is unsuppressed. In the following we present the bounds derived by various authors based on this process. 

\begin{table}
\begin{tabular}{cccccccc}
\hline
  & \tablehead{1}{c}{c}{}
  & \tablehead{1}{c}{c}{Cullen and \\Perelstein\\ \cite{Cullen:1999hc}}
  & \tablehead{1}{c}{c}{Barger,\\ {\it et. al.}\\ \cite{BHKZ}}
  & \tablehead{1}{c}{c}{Hanhart\\ {\it et. al.}\\ \cite{Hanhart:2001er}}
  & \tablehead{1}{c}{c}{Hanhart\\ {\it et. al.}\\ \cite{Hanhart:2001fx}}
  & \tablehead{1}{c}{c}{Hannestad\\ and Raffelt\\ \cite{Hannestad:2001jv}}
  & \tablehead{1}{c}{c}{Hannestad\\ and Raffelt\\ \cite{Hannestad:2003yd}} \\
\hline
$M_D$&n=2& 50 & 51  & 23.4  & 31   & 84 & 20.1\\
(TeV)&n=3&  4 & 3.6 & 1.51  & 2.75 &  7 & 1.26\\
\hline
$R_D$&n=2& $3 \times 10^{-4}$ &  $3.71 \times 10^{-4}$ & $7.1 \times 10^{-4}$ & $6.6 \times 10^{-4}$ & $0.9 \times 10^{-4}$ & $9.6 \times 10^{-4}$\\
(mm) &n=3& $4 \times 10^{-7}$ &  $5.39 \times 10^{-7}$ & $8.5 \times 10^{-7}$ & $8 \times 10^{-7}$ & $1.9 \times 10^{-7}$ & $11.4 \times 10^{-7}$\\
     \hline
\end{tabular}
\caption{Bounds from Nucleon-Nucleon Brehmstrahlung process in SN1987A}
\label{tab:a}
\end{table}

\section{{Conclusions}}

In summary, it has been found that KK graviton emission from SN1987A puts
very strong constraints on models with large extra dimensions in
the case $n=2$. In this case, for a conservative choice of the
core parameters we arrive at a bound on the $M_D \geq$ 30 TeV. 
We have done similar calculations in the case of plasmons which will be reported elsewhere. 
Even though taking into account various
uncertainties encountered in the calculation can weaken this
bound, it is unlikely that it can be pushed down to the
phenomenologically interesting range of a few TeV.
Therefore this case is still viable for solving the hierarchy problem and accessible to being tested at the LHC.

\subsection{Acknowledgments}
One of the authors (VHS) would like to thank the organisers of IWTHEP-2007 for their hospitality and for providing an excellent intellectual atmosphere. He thanks Professor R K Kaul for fruitful discussions and hosting at IMSc where part of this article was written under the Summer Research Fellowship of IASc, INSA and NASI. He also thanks Dr. R Chenraj Jain, Mr. K L Ganesh Sharma and Mr. V Venkatachalam for their encouragement.

\end{document}